\documentstyle[aps,prl]{revtex}
\begin{document}
\twocolumn

\narrowtext
{\large \bf Hohenberg-Kohn theorem is valid}
\ \\

In a recent paper, Gonze et al (GGG) \cite{1}, claim that the
Hohenberg-Kohn theorem (HKT), which is the basis of the widely used
density functional theory(DFT), is incorrect for periodic
solids in an external field, and propose an alternative
``density-polarization functional theory''. While the proposed
method itself is correct, and may become useful in practice,
 the statement about invalidity of the HKT is
wrong, and the useful technical point of Ref. \cite{1} is wrapped
into a set of formally incorrect and misleading statements.

First it is necessary to separate inaccuracies of the
local density approximation (LDA) and fundamental problems with the
DFT which are confused in GGG.  GGG state that the DFT
 fails to describe static dielectric response for
semiconductors. In reality, accurate DFT calculations
for dielectric response, even in the LDA, are very complicated,
because the dielectric function strongly depends on the local field
corrections in form of Umklapp processes and exchange-correlation (XC)
corrections. Existing calculations
give the results accurate within $\approx $10\% (GGG, Ref.5), which
is not bad, in view of the crudeness of LDA.  Note that
the XC interaction, $I_{xc}$, which defines the local
fields, is the {\it second }variation of the XC
energy, and thus very sensitive to the LDA.

The discontinuity of
of the XC potential, leadind to a difference between the
DFT band gap, $E_g^{DFT}$, and the real gap, $E_g^{exp}$, is the reason why
the DFT and the many body theory, which treat XC
prinicipally differently, must give the same static dielectric
function. To the contrary, GGG mention the difference between $E_g^{DFT}$
and $E_g^{exp}$ as an argument that DFT is not able to describe the static
response.

Taking this alleged failure of the exact DFT
for granted, the authors proceed with the
statement that the HK theorem does not apply to periodic solids in
an external field. Their argument is that an infinite solid in a
spatially unlimited homogeneous field does not have a ground state,
and thus the HK theorem, whose objects are ground states only, does
not apply.
This argument %applies as well to any variational
%treatment, not just DFT, and
is equivalent to the statement that
neither infinite solids nor unlimited fields exist in nature.
As soon as periodic boundary conditions are applied, this
argument fails.
 Rather what is necessary is to
use the method of long waves and treat a potential wave of
wavelength $q$, and one must make clear distinction between the wave
vectors $q<1/L$, and $q\gg 1/L$, where $L$ is the size of the
system\cite{DM} as one takes the infinite size limit. Since the HK theorem
applies to any physical system, like a macroscopically large
crystal, $L\gg a$ ($a$ is the lattice parameter), in a flat
capacitor with the plates much larger than $L$, or to a system
subject to an external field with finite $q$, it can be used to treat any real
physical problem.

The next misleading step in GGG is after making an argument
applicable only (with the reservation above) to the homogeneous
field, the authors consider a finite $q$ field, for which the
HK theorem holds without any reservation. What they
actually prove is a theorem which holds {\it in addition to}, but
not {\it instead of}, the HK theorem, namely that the total energy
of a periodic solid is a functional of the periodic density, $n_G=\sum_{\bf
G\neq 0}\exp((i{\bf (q+G)r})n({\bf q+G})$,
$n({\bf r})=\exp (i{\bf qr})n({\bf q})+n_G$,
 and macroscopic polarization. HK theorem states
that the same energy is a functional of the total density, including
$\exp (i{\bf qr})n({\bf q}),$ and this is correct.
  Eq. 8 in GGG which purports to
disprove HK, instead simply shows a separation of the long wave and
local parts of the problem, both of which must be included in HK
since HK deals with the whole system, not its parts.

The example given by GGG illustrates this: Using {\it periodic}
potential $\Delta V_{eff,2}$ they reproduced the correct
density within one cell, while the correct, {\it non-periodic}
potential $\Delta V_{eff,1}$ gives correct density for the whole supercell.
 There
is nothing counterintuitive in this result, nor is it ``in contrast to a
naive application of DFT''.

Finally,
discussion of the ``scissor correction'' is also misleading.
A scissor correction enlarges the gap, bringing the RPA polarizability
of the Kohn-Sham particles closer to the RPA polarizability of real
one-electron excitations. Fully renormalized polarizability, however,
includes the local field corrections (note that Eq. 11 of GGG is just the
long-wave part of the XC local field). The latter are  principally different
in the  DFT and in the many-body theory (e.g., they have qualitatively
different behaviour at ${\bf q+G\rightarrow 0}$). Of course, the fact
that $I_{xc}$ in LDA is a contact interaction, may constitute a large
difference between LDA and exact DFT, reflected in a difference of
 $E_g^{LDA}$ from $E_g^{DFT}$,
which probably may be corrected by a scissor-like technique. It should
not be mixed with the difference between $E_g^{DFT}$ and  $E_g^{exp}$,
which  should {\it not} be corrected at all, since this difference is to be
cancelled by the appropriate difference in the XC corrections.

%So, possible errors in the LDA gap,
%compared to exact DFT gap\cite{Jap} is not the only one, and maybe even not
%the main source of error in the LDA calculations of $\epsilon $, the error
%in $I_{xc}$, first of all incorrect functional form (in LDA it depends only
%on the difference ${\bf G-G}^{\prime }$, instead of separate dependence in
%DFT), being another one.

To conclude, the claim of Gonze et al that DFT is invalid
for dielectric response, and should be substituted by a density-polarization
functional theory, is incorrect, but the theorem they have proven (that the
total energy is a functional of polarization and periodic part of density)
is correct and may become useful in the future. We also would like to thank
J. Serene for discussion of this Comment.
\vskip 3mm
\noindent I.I. Mazin and R.E. Cohen

{\small Carnegie Institution of Washington

5251 Broad Branch Rd., Washington, DC 20015\\
30 May 1995}


\vskip -2.1cm
\begin{references}
\bibitem{1}  X. Gonze, Ph. Ghosez, and R.W. Godby, Phys. Rev. Lett.,
{\bf 74}, 4035 (1995).


\bibitem{DM}See, e.g.,
  O.V. Dolgov and E.G. Maksimov, in: {\it The dielectric function
of condensed systems}, Elsevier, 1989.

\end{references}
\end{document}